# DAG With Omitted Objects Displayed (DAGWOOD): A framework for revealing causal assumptions in DAGs

Noah A Haber, Mollie E Wood, Sarah Wieten, Alexander Breskin*

Noah A Haber, ScD
Meta Research Innovation Center at Stanford University
1265 Welch Rd
Palo Alto, CA 94305
(650) 497-0811
noahhaber@stanford.edu

Mollie E Wood, PhD
Department of Epidemiology
Harvard TH Chan School of Public Health
677 Huntington Avenue, Boston MA 02115
(617) 432-7918
mwood@hsph.harvard.edu
Funding: 1T32HL098048 from NHLBI

Sarah Wieten, PhD
Meta Research Innovation Center at Stanford University
1265 Welch Rd
Palo Alto, CA 94305
(650) 497-0811
wietens@stanford.edu

Alexander Breskin, PhD
NoviSci, Inc
PMB 218
201 W Main St, Ste 200
Durham, NC 27701
(917) 593-9004
abreskin@novisci.com

* Authorship is shared, no senior author position

# Abstract

Directed acyclic graphs (DAGs) are frequently used in epidemiology as a method to encode causal inference assumptions. We propose the DAGWOOD framework to bring many of those encoded assumptions to the forefront.

DAGWOOD combines a root DAG (the DAG in the proposed analysis) and a set of branch DAGs (alternative hidden assumptions to the root DAG). All branch DAGs share a common ruleset, and must 1) change the root DAG, 2) be a valid DAG, and either 3a) change the minimally sufficient adjustment set or 3b) change the number of frontdoor paths. Branch DAGs comprise a list of assumptions which must be justified as negligible. We define two types of branch DAGs: exclusion branch DAGs add a single- or bidirectional pathway between two nodes in the root DAG (e.g. direct pathways and colliders), while misdirection branch DAGs represent alternative pathways that could be drawn between objects (e.g., creating a collider by reversing the direction of causation for a controlled confounder).

The DAGWOOD framework 1) organizes causal model assumptions, 2) reinforces best DAG practices, 3) provides a framework for evaluation of causal models, and 4) can be used for generating causal models.

# Introduction

Directed acyclic graphs (DAGs) are emerging as one of the most important conceptual frameworks in epidemiology. They provide a formal graphical structure paired with a rule set for communicating and understanding causal relationships, and a corresponding formal calculus and structure for causal identification (1–3). In theory, they guide both creators and consumers of research toward deeper understanding of the causal model relevant to their problem by clearly describing the causal model. However, their hypothetical benefits do not necessarily translate well into application (4) or epistemic theory (5,6). In practice, DAGs call little attention to the causal assumptions they encode and provide little formal structure for prospectively generating and critiquing causal models.

Missing pathways (i.e. those whose magnitude is assumed to be equal to zero) and the direction of existing edges are two of the most critical causal assumptions for any causal model. The sharp causal null (2,7) and alternative edge direction assumptions are denoted by the *absence* of a pathway connecting two nodes and alternative pathways, respectively. Violations of these key hidden decision assumptions can lead to substantial issues in effect estimates, particularly where observational settings are the only practical option (8).

We propose a DAG-based framework to help identify and display those key hidden assumptions: DAG with omitted objects displayed (DAGWOOD). DAGWOODs take an existing root DAG, generate a set of alternative DAGs representing key assumptions, produce a corresponding overlay to display them over the original DAG, and list those alternative assumptions.

DAGWOOD is designed to assist assessing strength of causal inference by revealing and highlighting key underlying assumptions for a given analysis. It 1) makes explicit and organizes the most important causal model assumptions, 2) reinforces best DAG practices, 3) provides a framework for critical evaluation of causal models, and 4) provides an iterative process for generating causal models.

# DAGWOODs

## Conceptual overview

The DAGWOOD framework consists of four parts: the root DAG, a set of branch DAGs, the DAGWOOD overlay which graphically represents these branch DAGs, and a list that represents the same. Definitions are documented in Appendix 1 for reference.

The root DAG represents the causal model as implied by what is implemented *in analytical or statistical practice*. When researchers address a causal question with quantitative/statistical methods, a DAG is implied both by the final model selected for estimation and by the language used to describe the question and/or its implications (i.e. the implied causal estimand). This root DAG includes exposure(s), an outcome, and any covariates selected and adjusted for, but not, by definition, anything that was not adjusted for.

Root DAGs allow the DAGWOOD to explore how practice might differ from theory based on encoded assumptions. The root DAG may or may not be ideal or correct: there may be unmeasured and/or omitted confounders, incorrectly specified causal structure, missing pathways, and/or any combination of these, but it is the implemented (or proposed) DAG. If this (root) DAG does not pass muster, neither does the implemented statistical analysis with which it is associated.

The root DAG is constructed from the implied causal estimand, and contains the full set of nodes which are included and adjusted for in the statistical analysis. It must contain at least an exposure (A) and outcome (Y), and could include covariates and/or instruments. It does not contain any other nodes or edges beyond those included in the statistical adjustment set implemented or considered for implementation in practice.. In the DAGWOOD framework, those hypothetical (but not implemented) edges and nodes which may exist enter into the DAGWOOD separately as branch DAGs.

While the arrangement of nodes and arrows in the root DAG is set by the items included in the statistical model, their arrangement into the root DAG (i.e. the placement and direction of arrows) is set by the implied causal estimand. In some cases, there may be a provided DAG figure, but there is no guarantee that this figure represents the implied causal estimand as needed to design the root DAG for DAGWOOD. For example, such a DAG may be arranged such that it has a mediator adjusted-for when the implied causal estimand is a total effect or vice versa, it may include nodes not adjusted-for in the statistical analysis, and/or may have arrow arrangements otherwise not matching the implied causal estimand.

The process for constructing a root DAG will vary depending on the specific use case. In the case of reviewing an existing paper, one place to start is in the model specification section. Identify the exposure, outcome, and adjusted-for variables, and any text justification or discussion to help work backward to draw the DAG implied by those variables. The same can be done while DAG building alongside a dataset, where the user builds a DAG with respect to the available data and causal theory, and/or with the DAG building procedures described in a later section.

DAGWOODs comprise a set of branch DAGs which are modifications of the root DAG, and which are intended to represent omitted or hidden assumptions about the root DAG with respect to the effect of the exposure on the outcome. These branch DAGs are independent of each other, and exist only with respect to the root DAG.

DAG could be considered a branch DAG in the DAGWOOD framework if it passes the three conditions below:

1. Must modify the root DAG (i.e. change or add at least one edge), AND
2. The resulting edges and nodes must interpretable as a valid DAG (e.g. must be acyclic), AND
3. The resulting DAG must require or result in a change of either:
    a. The adjustment set to estimate the effect of interest from the root DAG, OR
    b. The number of frontdoor paths from the treatment to the outcome.

Condition 2 can be checked using any graphical algorithm for detecting cycles, such as a depth first search algorithm (9). Condition 3 can be checked using identification procedures such as that proposed by Schpitser and Pearl (10). Condition 3 makes condition 1 redundant, but aids in computation when implemented due to not needing to check the adjustment sets in these cases.

We introduce two types of branch DAGs: exclusion and misdirection branch DAGs. Exclusion branch DAGs represent additional causal pathways and elements assumed to be sharp causal nulls in the root DAG. Misdirection branch DAGs represent alternative DAGs which could be drawn with the same set of nodes and edges, but with one or more having reversed directions.

*Figure 1: Example root DAG, DAGWOOD overlay, and branch DAGs encoded in DAGWOOD overlay with one known omitted confounder*

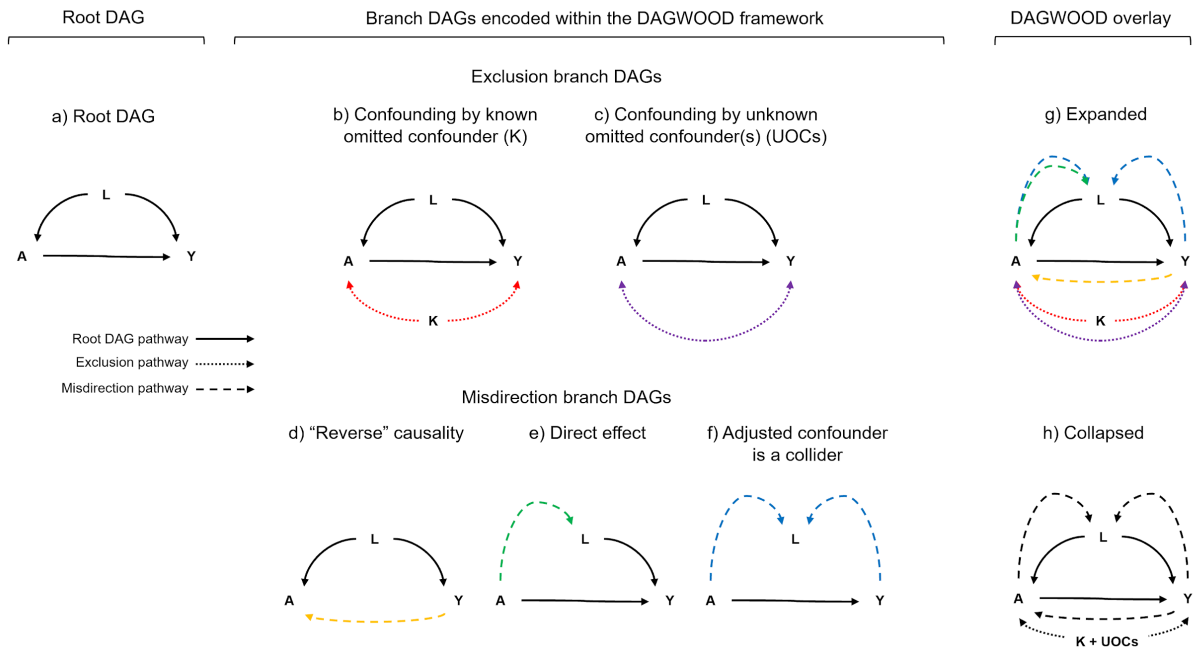

A is the exposure of interest, Y the outcome of interest, and L is a measured confounder, as well as one known omitted confounder, K. Displayed in this figure are: (a) the root DAG from the original or proposed analysis, branch DAGs (b) - (f), (g) the DAGWOOD overlay which combines the root and branch DAGs, and (h) the collapsed DAGWOOD overlay. The middle section shows the five branch DAGs generated from the root DAG plus the known omitted confounder. Branch DAG pathways are color coded to correspond to the pathways in the DAGWOOD overlay on the right.

Figure 1 shows the root DAG, the five branch DAGs within the DAGWOOD, and the DAGWOOD overlay for displaying this in one figure. The simple root DAG (a) with an exposure A, and outcome Y, and a measured confounder L generates five branch DAGs, each corresponding to a key assumption already contained within the root DAG. The two exclusion branch DAGs display additional confounding pathways from (b) a known but omitted confounder K, and (c) an unknown omitted confounder. The three misdirection branch DAGs show scenarios in which (d) the outcome causes the exposure, or the proposed confounder L is either (e) a mediator or (f) a collider (11).

Branch DAGs are compiled into a list corresponding to key assumptions that would be violated if the scenario in the branch DAG, rather than the root DAG, is correct. Each potential violation must be justified as negligible or impossible. This serves as a checklist, where each assumption should be addressed directly. If any one item in this list cannot be falsified, causal identification assumptions may be violated. This provides structured guidance and an iterative process to building and determining the completeness of the DAGs at any point in model building.

Branch DAGs may also be represented in the form of a graphical overlay, as shown in panels g) and h) in Figure 1. The DAGWOOD graphical overlay consists of the original root DAG plus the additional edges present in at least one branch DAG. These overlay edges represent the changes from each branch DAG resulting from the root DAG, and correspond directly with branch DAGs. The overlay is graphical shorthand for showing many related branch DAGs at once, but is not itself a DAG. In practice, this can be visually simplified by collapsing supersets (discussed in the subsequent section) and/or branch DAGs which share, where g) shows an expanded overlay and h) shows the same fully collapsed, an additional example of which is shown in Appendix 2.

Together, DAGWOODs take what is used in practice (the root DAG), identify alternative and hidden assumptions (branch DAGs), and highlight them (lists and graphical representations).

## Generating branch DAGs

### Exclusion branch DAGs

Exclusion branch DAGs describe causal pathways that are assumed to be negligible or non-existent. The terminology borrows from the term "exclusion restriction" in econometrics, which most commonly refers to the keystone assumption in instrumental variables (IVs) in which the instrument must have no conditional causal relationship with the outcome of interest except through the exposure (i.e. that a pathway violating this assumption does not exist) (3,12,13). The same logic applies to causal connections between unconnected nodes and/or missing common cause pathways (e.g. confounders). Broadly, assumptions in which a hypothetical causal pathway must not exist are assumptions of exclusion. The most common form of exclusion assumptions are those requiring an assumption of no confounding. The pathway in question is a common cause, or more informally a "bidirectional" pathway, in which two nodes in the root DAG are connected by an assumed node pointing to each of the two.

Exclusion branch DAGs represent mono- and bi-directional causal pathways that the implied causal model must assume are negligible to allow valid or unbiased estimation of the effect of interest. These pathways can be between any two existing nodes in the root DAG, whether already connected by an edge or not. While unmeasured confounding is likely to be the most common form of exclusion pathway, not all exclusion branch DAGs will represent confounding pathways.

To address known potential variables or mechanisms which violate exclusions in the root DAG, we make the distinction between unknown omitted pathways (generated automatically) and known omitted pathways (specified by an analyst). Making a distinction between the known and unknown pathways serves to document components that the analyst is already aware of and separate these from newly found concerns in preparation for model building, review, and evaluation. Because the automatically-generated exclusion branch DAGs represent a complete set of potentially violating pathways, the known omitted mechanism will typically match the edges of an existing unknown omitted exclusion branch DAG. We designate these "(un)known omitted pathways" rather than "(un)known unmeasured" to acknowledge that it is possible both to have measured a variable of interest but not included it in the implied causal model, and to know of the existence of a confounder but not have measured it.

Each exclusion branch DAG represents a superset of possible causal pathways. Each known omitted pathway may either be considered in its own exclusion branch DAG, or as a member nested within a pre-existing exclusion branch DAG. In the latter case, the superset contains multiple named members, including all known omitted and the unknown omitted member. The unknown omitted member of each exclusion pathway is itself a set of potentially infinite mechanisms, containing any possible unknown mechanism which is not covered by the root DAG or the known exclusion pathways. The unknown omitted member is the residual marginal causal effect remaining in an exclusion pathway, conditional on the root DAG pathways and all known members within that exclusion pathway superset.

Generating exclusion branch DAGs

The simplest algorithm for exclusion branch DAGs and pathways starts with listing each possible pair of nodes in the root DAG. For each pair of nodes, the algorithm independently evaluates three new edges: one direct edge in each direction between the nodes, and one bidirectional edge from an assumed unmeasured node between them.

Each of the three drawn pathways is considered a valid exclusion branch DAG if it passes the common ruleset (i.e. must be a valid DAG that is different than the root DAG, and must require either a change in the adjustment set or the number of frontdoor pathways).

*Figure 2: Example exclusion branch DAG diagram with a mediator and a known omitted confounder*

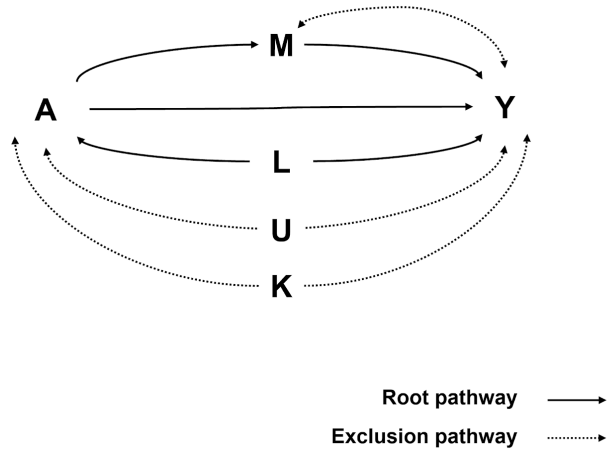

DAGWOOD overlay representing exclusion branch DAGs. The root DAG consists of an exposure (A), an outcome (Y), a measured confounder (L), and an adjusted-for mediator (M). In addition, there is a known omitted confounder (K). The overlay represents three branch DAGs, one for the known omitted confounder, K, a second for an unknown omitted bidirectional pathway (U) between A and Y, and a third for an unknown omitted bidirectional pathway between M and Y.

In Figure 2 we show a DAGWOOD overlay with exclusion branch DAGs with a mediator, M. Because M is adjusted for and on the causal pathway between A and Y, we are evaluating the controlled direct effect of A on Y, conditional on M. L and M are measured and adjusted for, since the root DAG includes only and all of the nodes adjusted for. Identifying the causal effect of A on Y means we assume Y = f(A, L, m, epsilon_Y) and A=f(L, epsilon_A). The causal direct effect of A on Y can be estimated by conditioning on L, and M.

To generate the exclusion branch DAGs, we have the following pairs of nodes to generate possible exclusion pathways:

    A-Y, A-L, A-M, L-M, L-Y, M-Y

Starting from A:Y, the A -> Y edge already exists and the A <- Y edge results in cyclicality, so neither is drawn as a DAGWOOD edge. The A<-U->Y and A<-K->Y pathways act as confounders in this case, and therefore require a different adjustment set than what was originally modelled to estimate the direct effect of A on Y. The DAG containing this bidirectional edge would therefore be added to the set of exclusion branch DAGs, where Y=f(A, L, K, epsilon_Y) and A=f(L, K, epsilon_A). A bidirectional edge between M and Y results in the DAG assumptions being violated without conditioning on the assumed latent variable, resulting in another exclusion branch DAG containing this bidirectional edge. No other pair of nodes results in an identified exclusion branch DAG. A bidirectional pathway between A and M, for example, does not induce a change in the required adjustment set or number of frontdoor paths, nor do any other combination of nodes in the root DAG, and therefore no other exclusion branch DAGs are present.

The known omitted exclusion pathway, K, may be considered a member of the set of exclusion containing this unknown omitted exclusion pathway, as it is also a bidirectional common cause pathway between A and Y. This could either be expressed as two separate branch DAGs with similar bidirectional edges (one for the known and unknown exclusions each) as above, or as one collapsed branch DAG containing each of these members.

*Figure 3: Example exclusion branch DAG diagram with an instrumental variable*

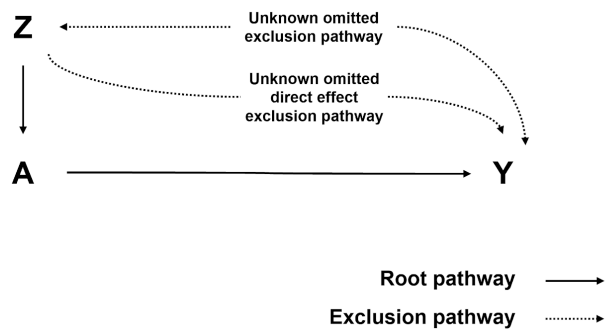

DAGWOOD overlay, representing exclusion branch DAGs with an exposure of interest (A), an outcome of interest (Y), and an instrumental variable (Z).

Figure 3 shows the exclusion branch DAGs associated with a typical instrumental variables model. This diagram has two independent branch DAG pathways corresponding with exclusion pathway violations. In the first, there is an uncontrolled common cause of Z (the instrumental variable) and Y. In the second, there is a direct causal pathway between Z and Y. Either would violate key causal assumptions in our model. Notably, the IV case is one where there does not exist a branch DAG representing confounding between A and Y, as the existence of such a confounder does not violate key causal inference model validity.

## Misdirection branch DAGs

Misdirection branch DAGs represent scenarios in which given causal paths are specified in the wrong direction. There are many ways this can occur, including conditioning on a collider of the exposure and outcome as if it were a confounder, simultaneity/reverse causality through hidden time nodes, simple mistaken direction, conditioning on downstream effects, and/or in more complicated cases such as M-bias or butterfly bias (14). Hidden time nodes could represent scenarios where the data generation process includes feedback between nodes that changes over time, but is masked by the observed data structure (15). That is often the case for cross-sectional analyses, where the data are measured at a single occasion, but are the result of causality in multiple directions via hidden time nodes. In another example a proposed confounder is actually a mediator, resulting in a shift from the target estimand (the total effect) to another estimand (in this case, the controlled direct effect).

The set of misdirection branch DAGs represents the DAGs which require the minimum number of flips of adjacent edges, for each edge in the root DAG to be flipped. In most cases, this will be only one flip. In others (e.g. for turning a confounder into a collider) this will require flipping two or more adjacent edges.

Misdirection and reverse causation is often a threat with retrospectively reported data. With such data, the analyst often must make an assumption regarding the temporality of reported events. This assumption is encoded in the root DAG by the directionality of the arrows between nodes, and carries great importance as it can influence both the set of confounders needed to control confounding, the set of colliders that should not be controlled, and the effect being estimated itself.

### Generating misdirection branch DAGs

Misdirection branch DAGs are based on nodes already connected in the root DAG, using a depth-first recursive algorithm to identify the branch DAG(s) requiring the minimal number of changes to the root DAG. For each pair of nodes with an existing edge between them, an algorithm starts by flipping the edge, and applies the common DAGWOOD branch DAG rules. If that satisfies all three rules, then the algorithm counts the resulting DAG as a misdirection branch DAG and moves to the next pair. If no valid branch DAG is located by changing the direction of the initial edge, the algorithm independently flips the direction of the unflipped edges attached to the newly downstream node and tests each of these new DAGs against the same rules. The process continues recursively (i.e. put an edge down, flip it, and recurse it) until at least one valid branch DAG is located, for each original pair of connected nodes.

*Figure 4: Example misdirection branch DAGs diagram with a mediator and a confounder*

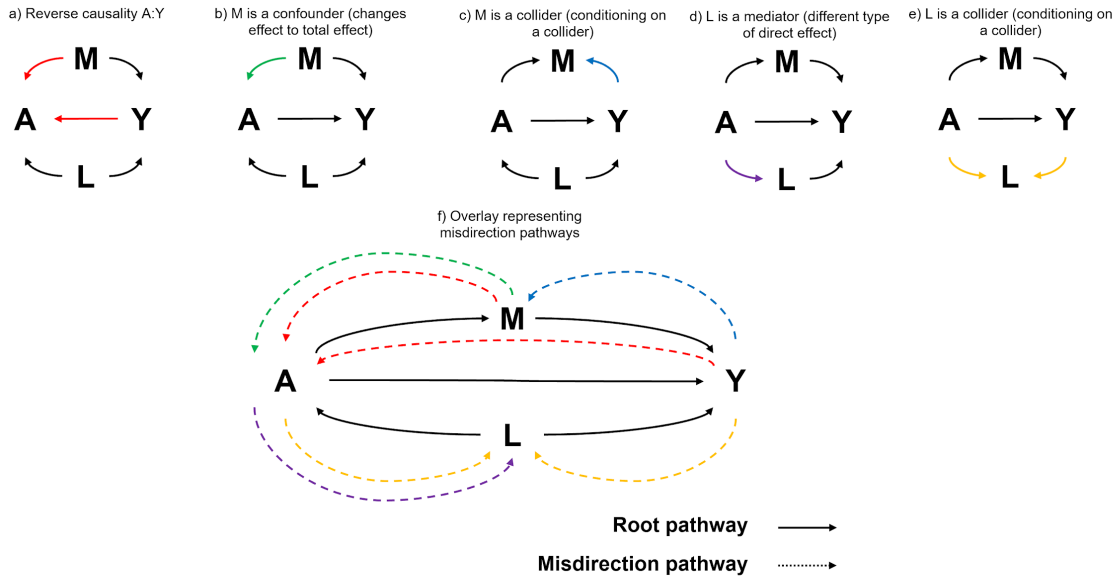

Example misdirections diagram with an exposure (A), outcome (Y), mediator (M), and confounder (L), color coded to link branch DAGs to DAGWOOD pathways. The root DAG is the same as in Figure 2.

Figure 4 shows the same root DAG as in the exclusion branch DAG section, but generating misdirection branch DAGs. Starting from the A->Y edge, reversing its direction alone would create a cycle due to the M, and therefore is not a valid misdirection branch DAG. The search continues from the A node, independently flipping the A->M and the L->A edges. In this case, the combination of flipping the A->Y + A->M passes, while A->Y + L->A does not, yielding panel a). Panels b) c) and d) require only a single flip, representing M being a confounder, M being a collider, and L being a mediator. Flipping the L->A edge on its own is a valid misdirection branch DAG representing what occurs when conditioning on a downstream effect. Flipping the L->Y edge does not pass, but the algorithm identifies also flipping the L->A edge where L is a collider, rather than a confounder.

Arrow directions may be fixed a priori such that the misdirection branch DAG will not flip pre-specified arrows, under the assumption and condition that the analyst already has sufficient justification that causality cannot run in that direction.

Importantly, a misdirection branch DAG that implies a change in estimand does not also imply that the shifted estimand is estimated without bias. Simply interpreting the misdirection branch DAG depicted in Figure 4d as a controlled direct effect is incorrect: analysts interested in the controlled direct effect should start their analysis from scratch and then go through the DAGWOOD algorithm for the revised analysis.

# DAGWOODs in practice

The DAGWOOD framework is designed to supplement and enhance DAGs, and implementation will necessarily vary slightly when using the framework as a pedagogical tool for introducing causal models, in the planning of data collection or analyses, or in critically reviewing research results. Interfaces can be built around any number of use cases to encourage engagement with the assumptions underlying causal models.

In one application, DAGWOOD can be used for model building. DAGWOODs with exclusion and misdirection branch DAGs collectively represent every possible minimal-change modification to a root DAG. All objects in the overlay represent an alternative DAG, and selecting one of those objects transforms the current DAG into that branch DAG. That in turn becomes the new root DAG, generating a new set of branch DAGs on which to expand and edit the model. At each iteration, the analyst would review the set of assumptions required for causal model validity. Where assumptions are unjustified, selecting a branch DAG directly corresponds to addressing that assumption. They can then iterate by running the DAGWOOD algorithm again until a satisfactory model is reached. Instead of drawing arcs and nodes from scratch, DAGWOODs let the user build a model through assumptions and conceptual threats to causal validity.

Another application is developing formal systematic causal inference review tools. A user would draw the DAG implied or presented by a given analysis, and the DAGWOOD algorithm highlights potentially vulnerable assumptions. While strength of causal inference review inherently relies on assessments from the reviewers (16,17), DAGWOODs can help make those assessments more systematic and replicable.

DAGWOODs are designed to be integrated into existing software packages for generating and displaying DAGs, such as DAGitty (18), to both generate and display DAGWOODs. A minimal implementation package for demonstration purposes in R is available on https://github.com/noahhaber/dagwood/, using DAGitty as a backbone.

# Discussion

Causal inference is hard, and relies on untestable assumptions (1,3,19–23). In a DAG, the critical assumptions are hidden in the space between nodes and edges. While it is impossible to map the space beyond our best theoretical understanding (6,24,25), DAGWOODs highlight those assumptions, both as a warning, and a structure to systematically justify assumptions with content-area knowledge (1,3,15,26). Unlike existing DAG augmentations (27–29) and methods for assessing the magnitude and risks of bias (30), the DAGWOOD framework identifies vulnerable points in proposed DAGs and organizes them graphically and in text.

The DAGWOOD framework is based entirely on assumptions already encoded in DAGs, and makes no changes or additions to the underlying assumption structure of causal inference. Instead, it shifts the burden of proof for the assumptions that were already there. Rather than presenting a model under the assumption that it is valid, it presents the model under the assumption that it is invalid, and requires the user to justify otherwise. The DAGWOOD framework encourages positive affirmation that the most important causal model assumptions are plausible, rather than passive acceptance that they are not.

Assumptions can be justified by any combination of theory, study design, and/or testable data. In some cases residual confounding may be bounded (31) and/or compared in magnitude against an alternative (30). In others, negative controls (32), sensitivity analyses, and triangulation with other methods (33) are appropriate. Further, branch DAGs may have testable implications that are different from the root DAG, such as differing conditional independencies, which can be tested with data.

Importantly, DAGWOODs cannot identify the "correct" DAG or quantitatively test its assumptions. Further, they do not in any meaningful way address whether the model is applicable or useful, particularly with regard to external validity (34). Justified DAGWOOD assumptions are necessary, but not sufficient conditions for valid causal inference.

The framework is further expandable to include any other causal inference assumption violations that can be represented using DAGs, including measurement error and selection biases.. These may be implemented as to-be-developed additional algorithms or specified by hand. DAGWOODs also share similar properties and limitations as partial ancestral graphs (35), and can be used with complementary DAG-based augmentations, such as the fast causal inference algorithm (29).

While there is often infinite depth to key model assumptions for causal inference (36), not all assumptions are equally threatening to model validity. Causal inference models are incredibly fragile, requiring only one substantial violation of assumptions to fail. However, causal inference problems and scenarios may be chosen to reduce the impact of exclusion and model misspecification errors. Study designs limiting the opportunity for misspecification, such as randomized trials and quasi-experimental designs, may be more plausible than designs that rely entirely on model-based adjustment with only the available data.

The language we have chosen for DAGWOODs may help unify some of the language and concepts perceived to be in tension between econometrics and epidemiological approaches to causal inference. Referring to key assumptions by lack of causal connection as "exclusions" helps make it clearer that exclusion restrictions, residual confounding, and other potential biases are more similar in nature than often perceived. We hope that using this language and displaying these assumptions in DAG form can demystify cross-disciplinary conceptual differences and similarities, building bridges across fields and expanding methodological toolboxes (19,37).

One objection to the DAGWOOD framework might be that the complexity of representing all of the assumptions generated by the branch DAGs is overwhelming or unwieldy. This is, in some sense, the point: DAGWOODs illustrate and codify the rapidly multiplying assumptions already encoded in the DAG. Adding complexity and adjustment to a DAG is not free, and those assumptions displayed are only ever as overwhelming as what is encoded in the original root DAG. A list may be more practical than a graph as complexity increases, but the breadth of the assumptions remains the same. DAGWOODs do not generate new complexity so much as show the complexity that was already there.

In some cases when the DAGWOOD results indicate that the effect of interest is not identified, researchers may be able to address threats to validity by revising the study design or data source and re-estimating; in others, sensitivity analyses and quantitative bias analysis may provide bounds. However, we expect that some analyses will show that the target estimand is not identifiable in any meaningful way given the data and methods available.

Becoming an "epidemiology of consequence" (38) requires that we confront the full weight of assumptions and limitations. DAGWOODs were designed to guide critical assessment of assumptions and limitations of our work at any stage from generation, review, meta-analysis, communication, and consumption. As Lesko et al., 2020 remark, "reciting identification assumptions like catechism or an incantation does not make them true. We must consider carefully whether they are met in each circumstance and design better studies to address instances in which they are not," (23). We hope that DAGWOODs help put our assumptions and limitations before our results and conclusions.

# Appendix 1: Terminology and abbreviation definitions

Bidirectional pathway: A causal pathway connecting two nodes, with an assumed common cause between them.

Branch DAG: A DAG generated in the DAGWOOD framework, representing exclusion and misdirection branch DAGs

DAG: Direct acyclic graph

DAGWOOD overlay: Graphical representation of the DAGs represented by exclusion and misdirection DAGs, generated from the DAGWOOD

DAGWOOD: Directed acyclic graph with omitted objects displayed. When alone, this refers to the combination diagram with the root DAG and its corresponding DAGWOOD overlay, which in turn represents the underlying branch DAGs

Edge: Causal connection between nodes in a DAG

Exclusion assumption: An assumption that no causal pathway exists between two DAG nodes, except through other nodes already accounted for in the root DAG

Exclusion branch DAG: A DAG that represents a scenario in which the exclusion assumption would be violated

Exclusion assumption pathway: A line that represents an exclusion branch DAG in the DAGWOOD overlay

Misdirection assumption: An assumption that the edge direction in a root DAG is not reversed

Misdirection branch DAG: A DAG that represents a scenario in which the misdirection assumption would be violated

Misdirection assumption pathway: A line that represents an misdirection branch DAG in the DAGWOOD overlay

Pathway: An edge or series of adjacent edges and nodes representing a causal connection in a DAG

Root DAG: The originating DAG on which the DAGWOOD is based.

Root DAG pathway: The graphical line that is drawn in a DAG or DAGWOOD representing an edge from the root DAG

# Appendix 2: Expanded and collapsed overlays

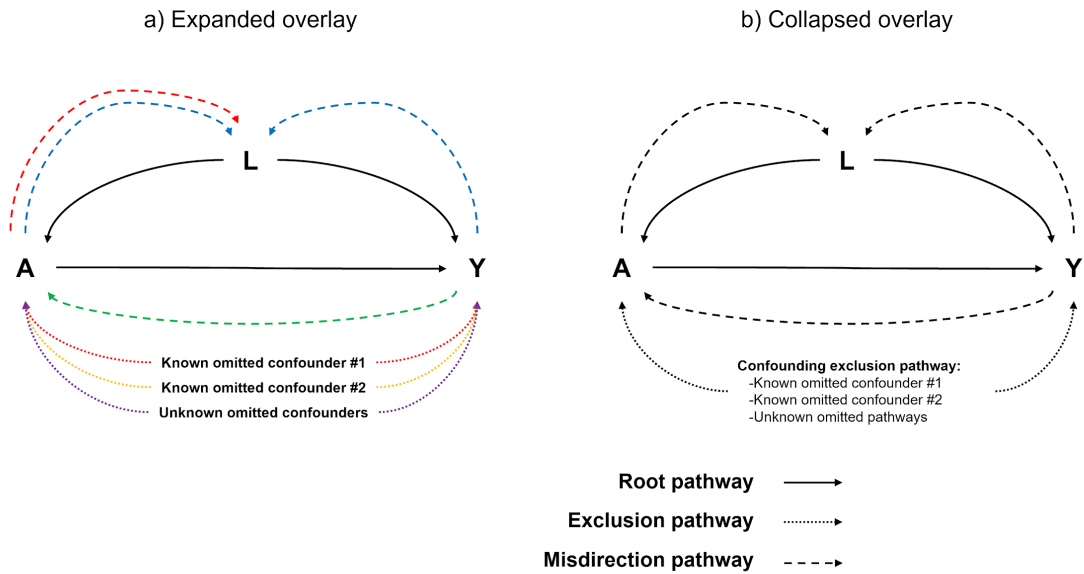

DAGWOOD overlay with an exposure (A), outcome (Y), a measured confounder (L), and two known omitted confounders. Panel a) shows the fully expanded overlay color coded to indicate each branch DAG, and b) shows the fully collapsed overlay.

# Acknowledgements

We would like to thank all of the people who provided critical commentary and suggestions on the many iterations of the design of DAGWOODs and manuscript draft, including Noam Finkelstein, Carlos Cinelli, Nick Brown, George Ellison, Johannes Textor, Maria Glymour, Robert Platt, Daniel Malinsky, Ilya Shpitser, Paul Zivich, Linnea Olsson, the Summer Epi Methods Journal Club, John Ioannidis, and Steven Goodman.

The authors declare that they have no financial or personal conflicts of interest to disclose. The Meta-Research Innovation Center at Stanford University is supported by Arnold Ventures LLC (Houston, Texas), formerly the Laura and John Arnold Foundation. Mollie Wood is funded through 1T32HL098048 from National Heart, Lung, and Blood Institute (NHLBI).